%% file: 0-index.tex
\documentclass[a4paper,conference]{IEEEtran}
\IEEEoverridecommandlockouts
\usepackage{cite}
\usepackage{amsmath,amssymb,amsfonts}
\usepackage{algorithmic}
\usepackage{graphicx}
\usepackage{textcomp}
\usepackage{xcolor}
\usepackage{cite}
\usepackage{amsmath,amssymb,amsfonts}
\usepackage{algorithmic}
\usepackage{graphicx}
\usepackage{textcomp}
\usepackage{amsmath}
\usepackage{subfigure}
\usepackage{epsfig}
\usepackage{color}
\usepackage{multirow}
\usepackage{caption}

\def\BibTeX{{\rm B\kern-.05em{\sc i\kern-.025em b}\kern-.08em
    T\kern-.1667em\lower.7ex\hbox{E}\kern-.125emX}}
\begin{document}

\title{Battery-assisted Electric Vehicle Charging: \\Data Driven Performance Analysis}

\makeatletter
\newcommand{\linebreakand}{%
  \end{@IEEEauthorhalign}
  \hfill\mbox{}\par
  \mbox{}\hfill\begin{@IEEEauthorhalign}
}
\makeatother

\author{
\IEEEauthorblockN{Junade Ali, Vladimir Dyo, Sijing Zhang}
\IEEEauthorblockA{\textit{School of Computer Science and Technology} \\
\textit{University of Bedfordshire}\\
 Luton, LU1 3JU, UK\\
vladimir.dyo@beds.ac.uk}
}

\maketitle

\begin{abstract}
As the number of electric vehicles rapidly increases, their peak demand on the grid becomes one of the major challenges.  
A battery-assisted charging concept has emerged recently, which allows to accumulate energy during off-peak hours and in-between charging sessions to boost-charge the vehicle at a higher rate than available from the grid. 
While prior research  focused on the design and implementation aspects of battery-assisted charging, its impact at large geographical scales remains largely unexplored. 
In this paper we analyse to which extent the  battery-assisted charging can replace high-speed chargers using a dataset of over 3 million EV charging sessions in both domestic and public setting in the UK. 
We first develop a discrete-event  EV charge model that takes into account battery capacity, grid supply capacity and power output among other parameters. 
We then run simulations to evaluate the battery-assisted charging performance in terms of delivered energy, charging time and parity with conventional high-speed chargers. 
The results indicate that in domestic settings battery-assisted charging provides  98\% performance parity of high-speed chargers  from a standard 3\,kW grid connection with a single battery pack.
For non-domestic settings, the battery-assisted chargers can provide  92\% and 99\% performance parity of high-speed chargers with 10 battery packs using 3\,kW and 7\,kW grid supply respectively. 
\end{abstract}


\section{Introduction}
\label{secIntro}
\input{intro.tex}

\section{Battery-assisted EV charging}
\label{secTechnology}
\input{concept.tex}

\section{System model}
\label{secModel}
\input{model.tex}

\section{EV charge point dataset}
\label{secDataset}
\input{dataset.tex}

\section{Experimental Results and Discussion}
\label{secResults}
\input{results.tex}

\section{Conclusions}
\label{secConclusion}
\input{conclusion.tex}

\bibliographystyle{IEEEtran}
\bibliography{references}

\end{document}

%% file: intro.tex
% !TEX root = 0-index.tex

Electric vehicles (EVs) represent a small but rapidly growing transportation segment due to lower carbon emissions, smaller running costs as well as higher energy efficiency. As battery costs decline, the EVs become more affordable and are set to replace traditional fossil-fuel based internal combustion engine (ICE) vehicles in the near future. 
At the same time, rapid charging of EVs remains a major challenge due to the need for high-powered grid connections in both domestic and urban settings \cite{Meyer2018ReviewIET}. 
Hence, the EV owners and charging station operators are required to upgrade their grid supply capacity,  which is  expensive to set up and increases the peak demand on electric grid. 

Recently, battery-assisted EV charging technology has been developed, where a large capacity battery is integrated into an EV charge point to accumulate energy during off-peak hours and in-between charging sessions and boost-charge the EV at a higher speed than is available from the grid. 
For example, a power-boost technology was developed by \cite{zapinamo}, which uses battery-stored energy to provide fast, rapid and ultra-fast charging from 3-7\,kW grid-supply. 
A system was developed by \cite{epfl}, which can charge a 30\,kWh car in 15 minutes using Li-ion battery bank using low power from the grid. 
The research led to a number of successful deployments in domestic and commercial sectors \cite{zapinamo} \cite{greenway}. 
At a national infrastructure scale, the UK National Grid plans to deploy a network of batteries across 45 battery sites near towns or major roads.
The \pounds 1.6 billion network will be used for rapid EV charging stations, support electric hub depots and will have capability to support 350 kW EV chargers. The system will connect directly to National Grid’s extra-high-voltage transmission system \cite{nationalgrid}.  

Prior work on battery-assisted charging addressed  the  design, performance evaluation of standalone chargers or its components \cite{AZIZ20171811} \cite{aziz2016battery}  \cite{Vasiladiotis2015}. 
In particular, Aziz et al \cite{aziz2017advanced} compare co-ordinated charging \cite{richardson2012local}, demand response \cite{deilami2011real} and battery-assisted charging technologies and  show  that battery-assisted charging  can enable fast charging whilst also minimising the stress experienced on the electrical grid that occurs when many electric vehicles are simultaneously charged. 
While prior work on battery-assisted charging investigated the design,  optimisation and performance analysis of individual charge points, its potential impact on charging performance at scale, taking into account the actual charge point usage requires further investigation. 

In this paper, we investigate the impact of battery-assisted charging on electric peak demand through a combination of simulation and a statistical analysis of a charge point usage dataset that contains more than 3 million charging sessions in domestic and urban settings in the UK.  
We first develop a formal model that describes a system state based on the battery capacity, power available from grid  and maximum charge output. 
The model is then driven by the charging events from the dataset  to understand the EV charging performance in terms of charging duration, dispensed energy and battery utilisation under domestic and public deployment scenarios for a range of energy storage capacities. 
To the best of our knowledge this is the first work analysing the performance of battery-assisted EV charging  at a large geographical scale.

%% file: concept.tex
% !TEX root = 0-index.tex

Depending on charging current, the Electric Vehicles charging speeds are categorised into slow, fast and rapid.
\begin{itemize}
\item Slow chargers connect to a standard 3-pin domestic socket. The standard UK socket is rated at 13\,A but normally provides 10\,A (2.3\,kW) to reduce the risk of overheating under continuous load. 
\item Fast chargers connect to a dedicated socket and deliver 7-22\,kW over a single or three phases. 
When using a fast charger, the charging speed will be limited by the power of vehicle's own onboard AC-to-DC convertor. 
Most cars on the market currently support single-phase max 16-32\,A charging (3.7-7.2\,kW) while the newer cars support charging with 2 or 3 phases (11\,kW and 22\,kW respectively).
\item Rapid chargers connect to a dedicated socket and deliver 43-120\,kW. 
Rapid chargers bypass the onboard charger of an electric vehicle and supply the DC current directly to the EV battery. 
DC charging stations are much more complex and more expensive than AC charging stations and require high powered electric supply point. 
\end{itemize}

\subsection{Battery-assisted charging system}

\begin{figure}[]
  \centering
   {\epsfig{file = 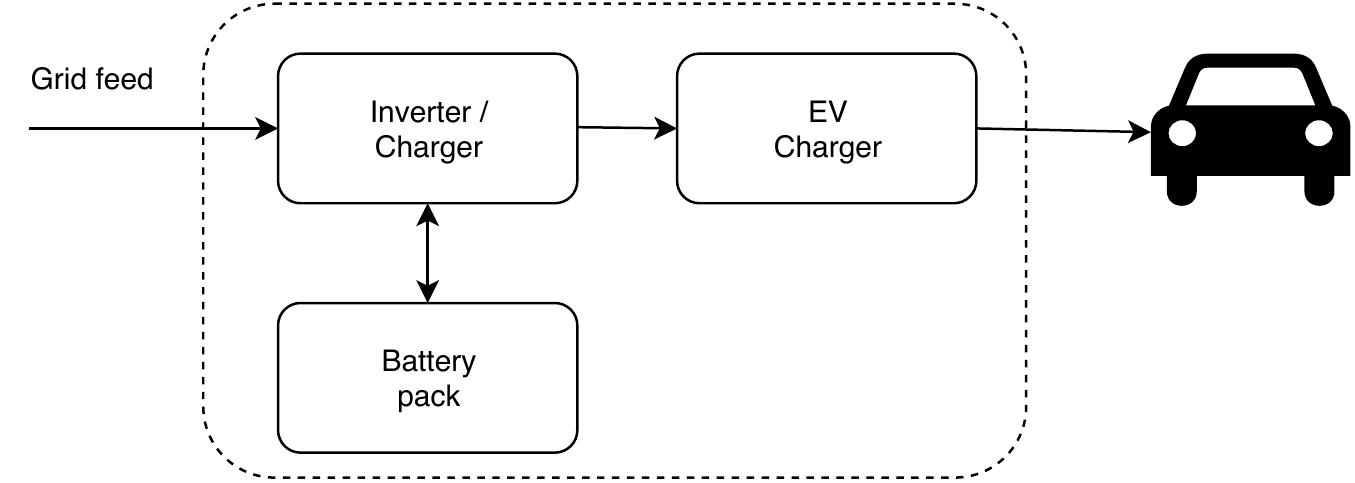, width = 7.5cm}}
  \caption{System Diagram.}
  \label{fig:architecture}
 \end{figure}

A battery-assisted charging system consists of a high capacity battery bank, a hybrid charging inverter and an EV charger as shown on Fig. \ref{fig:architecture}.  A brief description of each component is provided below. 
\subsubsection*{Battery} The battery accumulates energy from grid in-between EV charging sessions or when the EV charging speed is lower than available from the grid. 
The battery has capacity of $B_{storage}$ and provides a continuous output power $P_{battery}$ to the load. 
Several battery packs can be combined to increase capacity and the output current. 
\subsubsection*{Hybrid charging inverter}
This component converts the AC power to DC for   charging a storage battery, and then back to AC whenever battery is used to boost output power. 
The hybrid inverters with  pass-through capability have the ability to power the load from AC input while redirecting any excess to the battery. 
For example, Victron Quattro range of inverters which use battery-assist feature to combine AC mains and battery power  have been successfully used for EV charging systems \cite{victronQuattro}. 
\subsubsection*{EV charger}
EV charger communicates with the electric vehicle using a control pilot line to negotiate maximum allowable charging current, detect presence of the vehicle and control start and end of a charging session. 
The charging station can provide an output of $P_{battery} + P_{grid}$ unless the battery is depleted in which case the maximum output power reduces to $P_{grid}$. 
\subsubsection*{Electric Vehicle}
The maximum charging speed is limited by the vehicle's own AC/DC converter. 
Early EV models support single-phase  16-32\,A charging (3.7\,kW-7.2\,kW), while newer car models support 11\,kW and higher. 
The  charging duration depends linearly on the charging current and therefore can be very long on a standard 3\,kW supply. 
Table 1 compares approximate durations  to charge 40\,kWh battery using different charging speeds. 

\setlength{\tabcolsep}{4pt}
\begin{table}[]
\scriptsize
\caption{EV charging speed comparison}
\center
\begin{tabular}{ccccccccccc}
Supply type  / Charger rating &  AC/DC  &  Rated power &  Time to charge  \\
\hline
Regular socket, 10\,A 	& AC  & 2.3\,kW  &  17.4 hours  &     \\ 
Single phase, 32\,A 		&  AC & 7.3\,kW  &   5.5 hours &     \\ 
3 phase, 16\,A per phase 	& AC  & 11\,kW  &  3.6 hours  &     \\ 
 3 phase, 32\,A per phase 	& AC  &  22\,kW &   1.8 hours &     \\ 
 3 phase, 60\,A per phase	& AC  &  41\,kW &   58 minutes &     \\ 
3 phase, DC & DC  &  50\,kW &  48 minutes  &     \\ 
3 phase, DC &  DC  & 120\,kW  &  20 minutes  &     \\ 
\end{tabular}
\label{tab:resBulbDistribution}
\end{table}

%% file: model.tex
% !TEX root = 0-index.tex
\subsection{Charger model}
\label{model}
The EV storage battery state is described by the following system of equations: 
\setlength{\arraycolsep}{0.0em}
\begin{eqnarray}
&&B(t+1) = B(t) +  P_{gridfeed}\times T - EV(t) \\
&&0 \le B(t) \le B_{storage}\\
&&EV(t) \le   B(t) +  P_{gridfeed}\times T \\
&&EV(t) \le (P_{gridfeed}  + P_{battery}) \times T \\
&&EV(t) \le  P_{EVmax} \times T
\end{eqnarray}
\setlength{\arraycolsep}{5pt}
Where $B(t+1)$ is the amount energy stored in a battery at the start of timeslot $t+1$, which depends on the amount of energy $B(t)$,  energy received from grid $ P_{gridfeed}\times T$ and energy used for EV charging  $EV(t)$ in timeslot $t$. 
Here, we assume that the grid has a constant power capacity $P_{gridfeed}$. 
Constraint (2) reflects the maximum battery capacity, $B_{storage}$. 
Constraint (3) ensures that in any single timeslot, the amount of energy spent on EV charging does not exceed the combined energy from the grid and the battery. 
The power supplied by grid, $P_{gridfeed}$ can be 2.3\,kW for domestic chargers using UK 3-pin socket and ranges from 3\,kW to 100\,kW+ for dedicated EV charging stations. 
Constraint (4) limits the delivered power to a combined power of grid and a battery pack $P_{gridfeed}  + P_{battery}$, which  can be increased by connecting multiple batteries in parallel. 
Constraint (5) limits the EV charging speed to a maximum EV charging power capability, $P_{EVmax}$. 
$T$ is a timeslot duration, which without loss of generality is assumed to be short enough that it contains no more than a single charging session. 
The battery pack charges whenever an EV charger is idle or is supplying less than $P_{gridfeed}$  to EV.  
In the event that more energy is needed than the grid can supply, the charger's battery must be used. 
The maximum supplied power is also limited by the capacity of battery pack and the inverter.

%% file: dataset.tex
% !TEX root = 0-index.tex

The experimental evaluation is based on electronic charging station (ECS) usage datasets in the UK, provided by the Department of Transport's Energy and Environment statistics \cite{domesticdata2018}\cite{localauthoritydata2018}. 
Each dataset contains raw data on  amount of energy supplied, and plugin duration per charging event for both domestic charge points and non-domestic (local authority) settings in the UK. 
The data analysis has been performed in $R$ statistical environment \cite{R}. 
 
\subsection{Domestic charge point dataset}
The domestic charge point dataset contains 3.17 million charging sessions from 25126 domestic chargers collected in 2017  in the following format:
\begin{table}[!h]
\label{tab:domestic-format}
\tiny
\begin{tabular}{llllllll}
\hline
EventID	& CPID 		& StartDate	& StartTime	& EndDate	& EndTime	& Energy	& PluginDuration\\
3177742	&	AN21771	&31/12/2017	&23:59:23		&01/01/2018	&18:20:23		& 8.8		&18.35\\
16679268	&	AN04715	&31/12/2017	&23:59:00		&01/01/2018	&00:03:00		&10.2	&0.066\\
\end{tabular}
\end{table}

The  session charging speed, computed separately by dividing the dispensed energy by the plugin duration, varies within the same charge point as 
plugin duration can be longer than the actual charging duration. 
To eliminate this factor from our analysis we introduced an {\em effective charging duration} metric as ratio of dispensed energy to the maximum charging speed within the given EV charge point as described in Section \ref{secDomAnalysis}. 
Table \ref{tab:maximum-charging-events} shows the distribution of maximum charging speeds for various charge speed intervals, which can provide an idea about the number of EVs in each charging speed category.

\begin{table}[b]
\caption{Distribution of charging session speeds. Domestic chargers}
\label{tab:maximum-charging-events}
\scriptsize
\begin{tabular}{lll}
\hline
Session speed, E(kW)     & \#CPs with this Max rate & \#CPs with this median rate \\ \hline
0.0 $\leq$ E $<$ 1.0     	& 393	&	11572\\
1.0 $\leq$ E $<$ 2.3     	& 691	& 	8755\\
2.3 $\leq$ E $<$ 3.0     	& 1292	& 	1871\\
3.0 $\leq$ E $<$ 7.0    		& 15095	& 	2707\\
7.0 $\leq$ E $<$ 22.0    	& 5887	& 	186\\
22.0 $\leq$ E $<$ 100.0  	& 1348	& 	26\\
100.0 $\leq$ E  			& 420	& 	9\\
\end{tabular}
\end{table}

 \begin{figure}[]
 \centering
  \includegraphics[width=8cm]{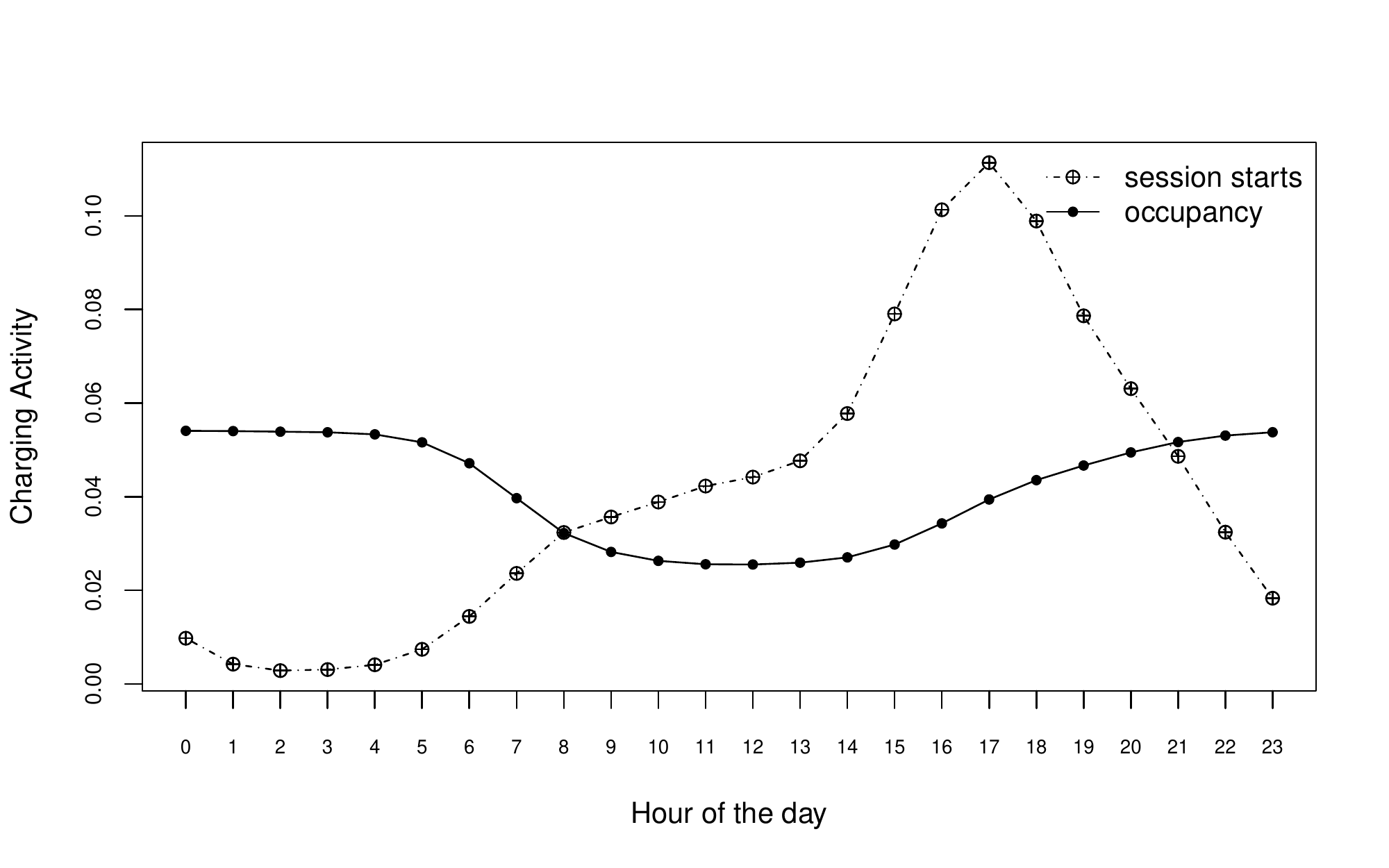}
  \caption{Domestic chargers diurnal activity distribution.  
} 
  \label{fig:domestic-sum-by-hour}
\end{figure}

 \begin{figure}[]
 \centering
  \includegraphics[width=8cm]{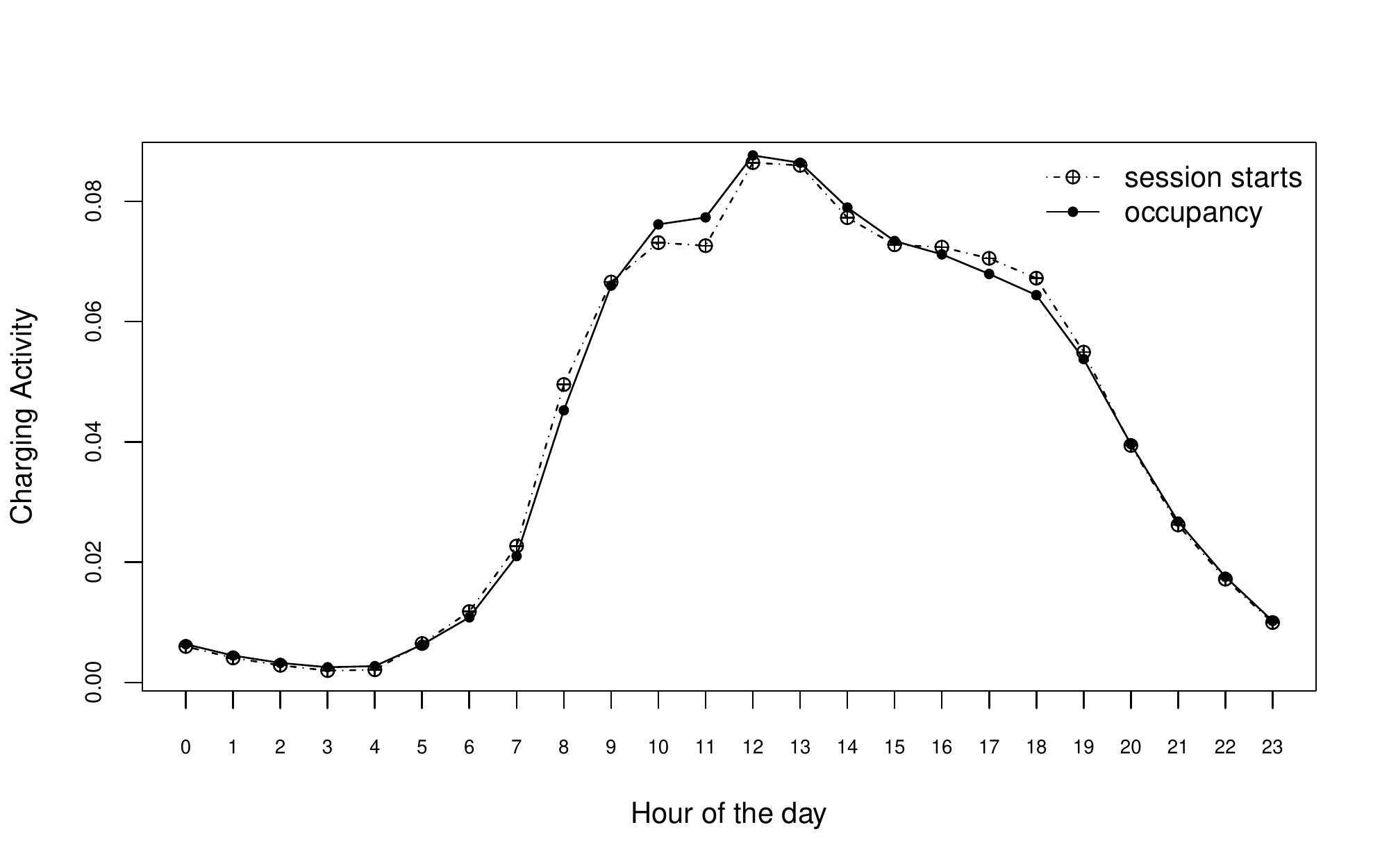}
  \caption{Local authority chargers diurnal activity distribution. }
  \label{fig:local-authority-sum-by-hour}
\end{figure}

For domestic charge point analysis, we consider chargers with a maximum charging rates between 3\,kW and 22\,kW, which represented 21009 charge points in total. 
As domestic chargers can typically charge only one EV at a time, the overlapping sessions within the same charge point, which represented 1.7\% of all charging events have been removed from the analysis.

Figure \ref{fig:domestic-sum-by-hour} shows that the charging session activity exhibits a strong temporal pattern.
The charging sessions are typically initiated between  1pm and 9pm with  vehicles remaining connected until early in the morning. 
As most sessions are relatively long in duration (the mean and median plugin durations are 12.44 and 10.72 hours respectively), 
the distribution  EV charger occupancy is more spread throughout a day compared to the distribution of  session start times.
The battery-assisted charging systems can accumulate energy during relatively quiet periods and release it during the peak hours to shorten the charge duration or reduce the peak load on the grid. 

\subsection{Local authority charge point dataset}

The local authority dataset contains 108746 charging sessions from 237 fast and rapid chargers deployed across $27$ local authorities in England \cite{localauthoritydata2018}. 
According to dataset, the vast majority of charge points are capable of drawing more than  22\,kW. 
For local authority charge point analysis, we consider chargers with a maximum charging rates between 3\,kW and 100\,kW, which represented 233 charge points in total. 
For the purpose of this study, we exclude charge points with multiple connectors and in cases where the number of connectors is not specified, assume that the charge point supports a single EV only. 
Table \ref{tab:speedDistributionAuthority} shows the distribution of charging and median speeds for public charge points.

\begin{table}[!h]
\caption{Distribution of charging session speeds. Local authority chargers}
\label{tab:speedDistributionAuthority}
\scriptsize
\begin{tabular}{c  c  c  }
\hline
Session speed, E(kW)     & \#CPs with this Max rate & \#CPs with this median rate \\ \hline
0.0 $\leq$ E $<$ 7.0    	& 2		& 	8\\
7.0 $\leq$ E $<$ 22.0    	& 1		& 	141\\
22.0 $\leq$ E $<$ 100.0  	& 230	& 	88\\
100.0 $\leq$ E  			& 4		& 	0\\
\end{tabular}
\end{table}

Fig \ref{fig:local-authority-sum-by-hour} shows the hourly distribution of dispensed energy and charge duration with most sessions initiated between approximately  7am and 11pm. 
Similarly to domestic charging, the behaviour of local authority chargers exhibits a strong temporal pattern although the peak usage is more spread throughout a day. 
Due to relatively short average charge session durations, both occupancy and session start distributions are  similar unlike those for domestic chargers (the mean and median plugin durations are 38.1 and 29 minutes respectively).

%% file: results.tex
% !TEX root = 0-index.tex

The goal of experiments is to evaluate the impact of battery-assisted scheme on charging performance in terms of mean delivered energy and charge duration when charge points are powered through standard 3\,kW grid supply. 
The evaluation shows an extent to which  battery-assisted charging can provide parity with existing rapid chargers  in terms of delivered energy.

The charge points are assumed to be equipped by a battery pack  with specifications similar to Tesla PowerWall 2 \cite{teslaPowerWall2}, which has  13.5\,kWh capacity and  5\,kW sustained power output  per battery.  
Several battery packs can be  combined in parallel to increase storage capacity and output power. 
We further assume that the EV can charge at a rate supplied by the charge point. 
The assumption is only used to estimate effective charging durations as explained later and does not affect the performance results in terms of amount of dispensed energy. 

\subsection{Domestic Charger Analysis}
\label{secDomAnalysis}
Table \ref{tab:domestic-comparison} shows the amount of delivered energy depending on the number of batteries. 
Interestingly, a charge point without any batteries can deliver 94.18\% of raw energy. 
This is because in domestic charging, the vehicle often remains connected to the charger for longer than the actual active charging duration and therefore reducing a grid supply  to  3\,kW does not significantly impact the amount of delivered energy. 
A single battery pack increases the proportion of delivered energy to 99.3\% while reducing the effective charging duration from 2.87 to 1.28 hours. 
Subsequent increase in number of batteries brings further improvements in effective duration but not in terms of energy delivered.  
In other words, given the amounts of dispensed energy and  relatively long plugin durations, downgrading the charge point grid connection to 3\,kW does not significantly impact the charging performance in terms of delivered energy. 
On the other hand, higher charging speed may encourage the drivers to charge more at home, as the median delivered energy is correlated with EV charge speed as shown in Table \ref{tab:dispensedVSspeed}.  

To estimate the impact on the actual charging duration, we obtain EV charging  capability as the maximum charging session speed within a given charge point, assuming it serves the same vehicle throughout a year.
We then define  {\em effective charging duration} as the ratio  of delivered energy to the EV charging speed. 
As can be seen in Table \ref{tab:domestic-comparison}, the effective duration for zero batteries scenario is 2.87 hours. 
As the number of batteries increases, the effective duration reduces to 0.42 hour due to higher charging speed provided by the charge point. 
Finally, the session parity rate shows a proportion of charging sessions that received the same amount of energy as in raw charging. 
As can be seen in Table \ref{tab:domestic-comparison}, the session parity rate ranges from 98.89 to 99.77 for a single and 4 batteries respectively. 

\begin{table}[t]
\scriptsize
\caption{Performance of Domestic Chargers}
\label{tab:domestic-comparison}
\setlength \tabcolsep{3.0pt}
\begin{tabular}{c  c   c c   c }
Batteries & Energy deliv.,kWh & Energy deliv.,\% & Effective duration,hrs & Parity,\% \\ \hline
\hline
 0 	&	8.60 & 	94.18 & 2.87 	& 	81.44\\
 1	& 	9.07 & 	99.30&  1.27	& 	98.89\\
 2	& 	9.11& 	99.82&  0.77	& 	99.61\\
 3	& 	9.12& 	99.91&  0.54	& 	99.75\\
 4	& 	9.12& 	99.92&  0.42	& 	99.77\\
\end{tabular}
\end{table}
\begin{table}[t]
\scriptsize
\caption{Distribution of dispensed energy by max charging speed. Domestic chargers.}
\label{tab:dispensedVSspeed}
\centering
\setlength \tabcolsep{3.0pt}
\begin{tabular}{c p{1.5cm} c   c }
\hline
EV Charging Speed range, kW & Energy (median),kWh & Energy (mean), kWh & \\ \hline
 0..3 	&	6.4	 &	6.97 	  \\
 3..7 	&	7.0	 &	8.10 	  \\
 7..11 &	9.7	 &	12.23  \\
11.22 & 	8.0	& 	9.96 \\
\end{tabular}
\end{table}

{\em Battery utilisation:}
The battery utilisation in terms of charge-discharge cycles is computed as  the total  battery discharge for all sessions within each charge point and dividing that by the battery pack capacity. 
The minimum, maximum and average battery cycles are 0, 482 and 49 respectively for a single battery. 
The actual utilisation may be higher as a battery may discharge and then charge within the same plugin session, however  the results  provide  a  baseline for battery life estimation.

\subsection{Local Authority Charger Analysis}
For local authority chargers we evaluate sessions with maximum charging speeds ranging from  3 to 100\,kW, which represent 98.7\% of all charging sessions. 
Table \ref{tab:authority-comparison} compares mean delivered energy, mean effective duration and parity rate for local authority chargers.  
The   results indicate that 3\,kW grid supply even in combination with high number of batteries is not sufficient to replace high-speed grid connections. 
\begin{table}[!h]
\scriptsize
\caption{Performance of Local Authority chargers}
\label{tab:authority-comparison}
\setlength \tabcolsep{2.0pt}
\begin{tabular}{l p{1.cm} l p{1.cm}  l p{1.cm} l  p{1.cm} l  p{1.cm} l  p{1.cm} l  p{1.cm} l  p{1.cm} }
Batteries &  \multicolumn{2}{c}{Energy deliv.,kWh} 	&  \multicolumn{2}{c}{Energy deliv.,\%} &  \multicolumn{2}{c}{Effec. dur,hrs} & \multicolumn{2}{c}{Parity,\%}\\
\hline
\hline
	 & 3kW 		& 7kW	 & 3kW 		&7kW	& 3kW 			 & 7kW 			& 3kW   				& 7kW \\ 
\hline
0	& 	1.79	& 	3.8	& 	15.57	&	33.12	&	0.59		& 	0.541		&	 3.62				& 	12.82\\
1	& 	4.21	& 	5.94	&	36.64	&	51.8	&	0.53		&	0.498		&	14.44			& 	22.96\\
2	& 	6.22 	& 	7.67	&	54.16	&	66.9	&	0.49		&	0.454		&	25.03			& 	39.09\\
3	& 	7.78	& 	8.97	&	67.71	&	78.2	&	0.45		&	0.411		&	41.04			& 	52.51\\
4	& 	8.88	& 	9.92	&	77.33	&	86.5	& 	0.41		&	0.370		&	53.44			& 	63.75\\
5	& 	9.63	& 10.57	&	83.85	&	92.1	&	0.37		&	0.334		&	63.15			& 	73.30\\
6	& 	10.12	&  11.06	&	88.15	&	96.4	&	0.34		&	0.302		&	71.20			& 	80.67\\
7	& 	10.44	& 11.36	&	90.93	&	99.00&	0.31		&	0.273		&	77.52			& 	90.77\\
8	& 	10.62	& 11.43	&	92.53	&	99.6	&	0.29		&	0.245		&	88.22			& 	99.45\\
9	& 	10.66	& 11.44	&	92.87	&	99.7	&	0.26		&	0.222		&	92.31			& 	99.57\\
10	& 	10.68	&11.44	& 	93.06	&	99.7	&	0.24		&	0.203		&	92.54			& 	99.63\\
\end{tabular}
\end{table}
\begin{table}[!h]
\scriptsize
\caption{Distribution of dispensed energy by max charging speed. Local authority chargers}
\label{tab:dispensedVSspeedAuth}
\centering
\setlength \tabcolsep{6.0pt}
\begin{tabular}{c p{1.5cm} c p{1.cm}  c p{1.cm} }
\hline
EV Charging Speed, kW & Energy(median), kWh & Energy(mean), kWh & \\ \hline
 0..11 	&	3.95	 &	5.91 	  \\
 11..22 	&	5.70	 &	5.35 	  \\
 22..50 	&	9.48	 &	11.17  \\
50... 		& 	8.34	& 	9.68 \\
\end{tabular}
\end{table}
With a single  battery, the battery-assisted charger can deliver only 36.64\% of energy achieving parity for only 14.44\% sessions.
As the number of batteries is increased to 10, the average delivered energy reaches  93.06\% with 92.54\% sessions achieving parity. 
The average effective duration with one battery is 0.5331 hours, which is  2.09 times slower than that of raw charging; as the number of batteries is increased to 10, the average effective duration approaches 0.2383. 
Increasing the electric supply to  7\,kW allows to achieve parity for 99.45\% of sessions with 8 batteries per charge point. 
Similarly to domestic chargers, higher capacity charge points dispense on average more energy, Table \ref{tab:dispensedVSspeedAuth}. 

 \begin{figure}[t]
 \centering
  \includegraphics[width=8cm]{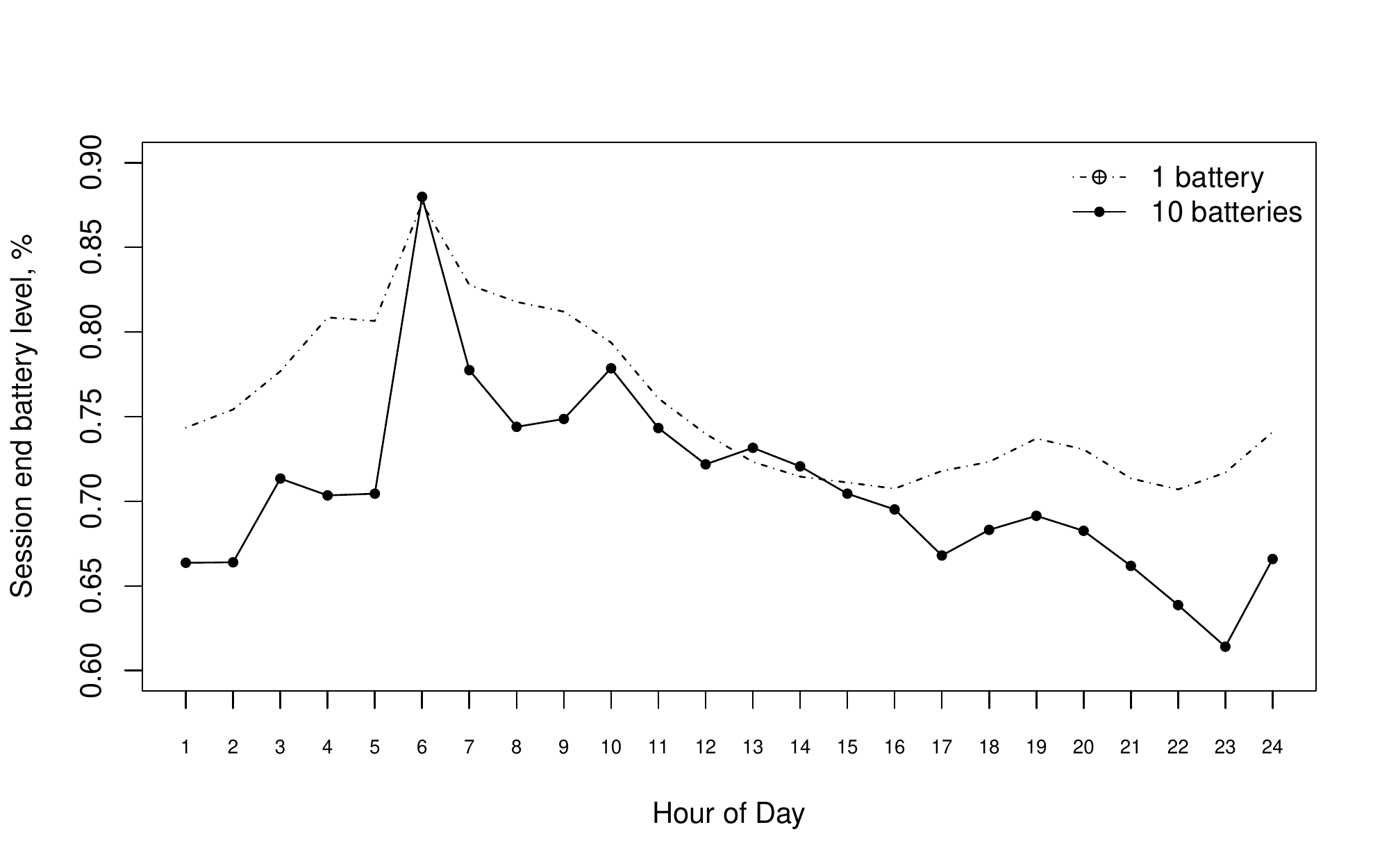}
  \caption{Mean battery charge levels by charging session start times for 1 and 10 batteries respectively. Local authority charge points with 3\,kW supply.}
  \label{fig:local-authority-bastarts}
\end{figure}

{\em Battery utilisation:}
Fig \ref{fig:local-authority-bastarts} shows  average battery charge levels for 1 and 10 batteries respectively.  
Each point on a graph shows a mean battery charge level for all sessions that ended at that particular hour of a day.  
Interestingly, the mean session end battery level for a single battery pack is higher than for 10 batteries. 
The explanation is that for the same amount of battery level (in terms of energy), a larger battery will discharge quicker  due to higher power output. 
The battery pack was fully drained only for 1.7\%, 5.9\% and 7.5\% charging sessions for 1, 5 and 10 batteries respectively. 
The analysis shows that the battery utilisation ranges from 0.08 to 491 cycles with a mean of 62 for a single battery pack, 
whereas the utilisation for a 10 battery pack ranges from 0.012 to 171 cycles with a mean of 24. 
Increasing the battery capacity reduces the  wear as expected, and the values obtained through simulation can provide a useful insight into the expected longevity and costs of  battery-assisted system.

%% file: conclusion.tex
% !TEX root = 0-index.tex

Electric vehicles have gained popularity recently due to their environmental benefits and energy efficiency, with governmental policies creating incentives to drive their ownership rates.
As the charging rate of a vehicle increases, so does the demand for the peak power and supply capacity of electricity grid  \cite{farkas2013grid}.
Battery-assisted charging  can facilitate the adoption of electric vehicles through lowering the cost barrier to introducing high-speed charging. 
In this paper, we have developed a  model for battery-assisted EV charge point and evaluated its performance against two publicly available charge point usage datasets in terms of delivered energy, effective charge duration and battery utilisation. 
Using simulations we analysed how battery-assisted charging technology can both compliment and act as replacements to high-speed chargers. 

The results demonstrate that in 98.89\% of domestic charging sessions, battery-assisted chargers were able to yield complete performance parity with domestic high-speed chargers (those rated between 3\,kW and 22\,kW) with just a single 13.5\,kWh battery using a standard 3\,kW grid supply.  
For local authority high-speed chargers, we have shown that a combination  7\,kW  supply with 108\,kWh battery can achieve  parity in terms of delivered energy for 99.45\% of all charging sessions. 
The renewable energy sources such as photovoltaics or wind generators can be used to further reduce the peak demand on the grid. 
Integrating these factors into analysis  is a potential future work.